\documentclass[final,5p,times]{elsarticle}

\usepackage{graphicx}
\usepackage{epsfig}
\usepackage{amssymb}

\usepackage{lineno}

\journal{Nuclear Instruments and Methods A}

\usepackage{url}

\begin{document}

\makeatletter
\def\@cite#1{[{#1}]}
\makeatother

\begin{frontmatter}

\title{Novel methods for measuring the optical parameters of the water Cherenkov detector}

\author[IHEP]{Xiurong Li\corref{cor1}}
\cortext[cor1]{Corresponding author. Tel: +86 10 88236035; Fax: +86
10 88236124} \ead{lixr@ihep.ac.cn}
\author[IHEP]{Huihai He}
\author[IHEP]{Gang Xiao}
\author[IHEP]{Xiong Zuo}
\author[IHEP]{Shaohui Feng}
\author[IHEP]{Lingyu Wang}
\author[IHEP,UCAS]{Cong Li}
\author[IHEP,UCAS]{Mohsin Saeed}
\author[IHEP]{Ning Cheng}
\author{for the LHAASO Collaboration}

\address[IHEP]{Key Laboratory of Particle Astrophysics, Institute of High Energy Physics, Chinese Academy of Sciences, Beijing, 100049, China}
\address[UCAS] {University of Chinese Academy of Sciences, Beijing, 100049, China}

\begin{abstract}
In this paper, novel methods are presented to measure the optical properties of muon detectors (MDs) in the Large High-Altitude Air Shower Observatory (LHAASO), which can also be used by other experiments. Each MD consists of a cylindrical water Cherenkov detector, with a Tyvek liner containing pure water, and a photomultiplier tube (PMT) mounted on the top. The time distribution of the photons collected by the PMT in the water Cherenkov detector have an approximately exponential distribution, and their decay factor is determined by the photon absorption length in the water, the reflectivity of the inner Tyvek surface, the mean reflecting step length of photons reflected by the inner surface, and the ratio of reflectional Tyvek area to the total inner surface area. By considering the principles of photon propagation in the water Cherenkov detector, we have developed novel methods to measure the water absorption length, the Tyvek reflectivity, and the mean step length of photons. The step length of photons can be determined by measuring the time distribution of the reflected photons hitting the PMT after multiple photons with single wavelength are generated in the tank, with slightly different ratios of the inner Tyvek to the total inner area. The water absorption length and Tyvek reflectivity can be measured simultaneously by the PMT by changing the height of the water, while the step length and Tyvek reflectivity in air can also be measured simultaneously. The proposed novel methods are supported by deduction of formulae and verified by GEANT4 simulations and the prototype experiment.
\end{abstract}


\end{frontmatter}


\section{Introduction}
Water Cherenkov detectors, containing pure water enclosed in a bag with Tyvek reflective inner surface, are widely used by several experiments, such as Pierre Auger \cite{auger}, the Daya Bay reactor \cite{dayabay}, Super-Kamiokande \cite{superk}, Tibet As array \cite{asgam} and the Large High-Altitude Air Shower Observatory (LHAASO) \cite{CAOZ_2014nim}. LHAASO is designed to search for cosmic ray sources by the measurement of gamma rays, and to study cosmic ray physics ranging from 10~TeV to a few hundreds of PeV \cite{CAOZ_2014}. The intended use of the 1171~water Cherenkov Muon Detectors (MDs) of the LHAASO, is to discriminate gamma rays from charged cosmic rays by detecting the muon content in the cosmic ray showers. For the LHAASO, each MD consists of a 6.8~m-diameter and 1.2~m-high concrete tank containing a sealed liner with a reflective inner Tyvek surface, and 44~t of pure water, and an 8-inch photomultiplier tube (PMT) mounted on the top to collect photons. Two MD prototypes of the same design, have been constructed at the Yangbajing Cosmic Ray Observatory (4300~m a.s.l.). Data analysis \cite{ZUO_2015} and detailed GEANT4 simulation \cite{lixr_2015} have been performed to understand the signals of the MD prototypes, and the theory of photon propagation has fully been studied. When a charged particle passes through the water Cherenkov detector faster than the phase velocity of light in water, it generates approximately 300~Cherenkov photons per cm with wavelengths in the range of 300--600~nm. A part of the Cherenkov photons propagate through the water, while a part of them absorbed or scattered by the water, a part of them absorbed or reflected by Tyvek, and a part of them hits the PMT to generate signals. The GEANT4 simulation, using the parameters of the MD as input, can help to understand the detector performance and reconstruct the number of muons. Thus, the measurement of the optical parameters of the water Cherenkov detector, such as the water absorption length, reflectivity of the inner Tyvek, and the Tyvek-reflection model is very important. These parameters are also important for understanding the optical mechanisms, which are measured and studied by several experiments with different methods.

The number of photons $N_{0}$ decreases to $N_{0} \times e^{(-\frac{x}{\lambda})} \times e^{(\frac{-x}{\lambda_{s}})}$ with the straight transmission distance of $x$, where $\lambda$ is the water absorption length and $\lambda_{s}$ is the water scattering length. The absorption length of water depends on the water purity and the wavelength of photons, which is difficult to be measured accurately for the pure water with long absorption length. The absorption length of pure water measured by different experiments are ranging from tens of meters to hundreds of meters with large inconsistencies, which is probably due to a combination of the experimental error and the different water purity \cite{length1}. Usually the attenuation length which is the combined effect of the water absorption and scattering effect is obtained by measuring the light intensity for photons passing through different distance of water \cite{length2}\cite{length3}, and then the absorption length can be calculated by subtracting the scattering effect. There are also certain methods to measure water absorption length effectively and independently of scattering effects at levels, with the results depending on the water quality in the equipment \cite{length4}\cite{length5}.

Tyvek is a paper-like material made with randomly distributed fibers of high-density polyethylene, which has high-reflectivity with diffuse reflection as the main part.
The water absorption length and Tyvek reflectivity depend on the wavelength of the photon, and reaches the maximum value with the wavelength about $\sim$400~nm. The maximum reflectivity of Tyvek in the water can be bigger than 98.5\%, which is greater than that in air. The reflectivity of diffuse materials is usually measured with the integrating sphere method, with a special and expensive equipment in the Bureau of Measures. The reflectivity of Tyvek 1082D used by LHAASO-MD prototype for photons with 405~nm wavelength in air, measured in the Bureau of and Measures is 93.9\%, with 1\% of uncertainty. The reflection model, related to the angular distribution of photons reflected by the Tyvek, may also be different for water as compared to air. Several experiments have measured the angular distributions of photons reflected by the Tyvek in certain directions \cite{angle1}\cite{angle3}. For the LHAASO muon prototype simulation, we used the UNIFIED Model to describe the Tyvek optical model using a ratio of diffuse reflection of $\sim$85\%, and others described as diffuse mirror reflection with a Gaussian-like angular distribution with a sigma of approximately 15$^\circ$. These parameters for the Tyvek model are similar to the parameters used by other experiments \cite{wly}. The mean transmission distance of photons between each reflection in the cylindrical water Cherenkov detector (the mean step length), is found to be constant after some number of reflections when there is enough photons in the tank. The mean step length depends on the Tyvek reflection model and geometry of the detector, is also a decisive factor for the performance of the detector. Cavity ring-down method is widely used by measuring the attenuation of photons reflected by two mirrors with fixed reflection optical path between each other \cite{cavity1}\cite{cavity2}. We found the optical parameters can be obtained by measuring the attenuation of photons in the cylindrical detectors with high-reflectivity and diffuse reflection dominated materials as the inner surface, because the mean step length of multiple photons in the tank becomes stable after several reflections.

We have proposed new methods to measure the water absorption length, the reflectivity of Tyvek, and the mean step length of photons based on the theory and experiment study, by using only the PMT to collect the time distribution of photons hitting it, while a pulse generator-driven LED provides single wave photons in the tank. These methods are supported by the deduction of formulae, and also verified by the GEANT4 simulation accompanied by the prototype experiment. The reflectivity for high-reflectivity materials with diffuse reflection as the main part can be measured by the new methods with bias less than 0.2\%. The mean step length of photons in the tank measured by the new methods is quite consistent with the values read out from simulation with high accuracy. The water absorption length can also be measured simultaneously, with accuracy higher than the long tube method. The water absorption length, the reflectivity of inner surface, and the mean step length of photons in the water Cherenkov detector can be measured directly by the detector, without the errors and contamination caused by the other equipments.

\section{Theory and methods}
\begin{figure}[!ht]
  \centering
 {\includegraphics[width=3.0in]{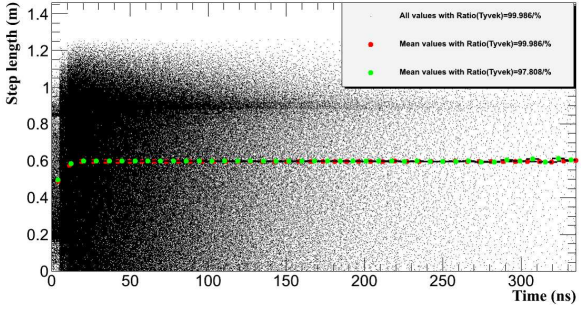}}
  \caption{Black points represent the simulated step length of photons at different time in the 90-cm diameter and 88-cm high small tank. The red points are the mean step length of photons in the small tank with its complete inner surface covered by Tyvek, except for the PMT area. The green points are the mean step length of photons in the small tank when the side Tyvek is shielded with vertical black strips to reduce the ratio of reflectional Tyvek area to approximately 97.8\%.}
  \label{mcsp_onestep}
 \end{figure}

\begin{figure}[!ht]
  \centering
 {\includegraphics[width=3.4in]{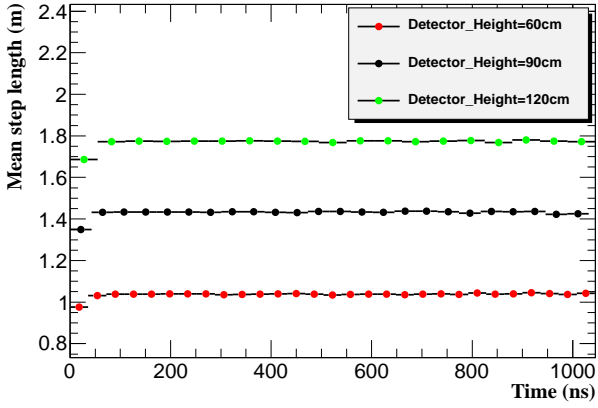}}
  \caption{Mean step length of photons in the LHAASO-MD prototype, averaged with the read out values from the LHAASO-MD prototype simulation with the 6.8-m diameter tank and different heights, with water absorption length of 100 m and Tyvek reflectivity of 98.5\%.}
  \label{muon_onestep}
 \end{figure}

The directions of photons in the cylindrical water Cherenkov detector are completely randomized after tens reflections by the Tyvek, and the mean step length of photons between each reflection becomes stable (see Fig. \ref{mcsp_onestep} and Fig. \ref{muon_onestep}). The transmission distance of photons between each reflection can be read out and supplied by GEANT4 simulation, which can be averaged to get the mean step length of photons in the tank. The black points in Fig. \ref{mcsp_onestep} represent the simulated step length of photons at different time after multiple photons is generated in 10 ns in a small tank with a diameter of 90~cm and a height of 88~cm, with a water absorption length of 100~m and a Tyvek reflectivity of 98\%. The red points in Fig. \ref{mcsp_onestep} are the profiled mean values of the black points, which are the mean step length for the small tank with the total inner surface of the tank is covered by the Tyvek except for a small hole for the PMT, while the green points are the mean step length of photons in the small tank when the side Tyvek is shielded with vertical black strips to reduce the ratio of Tyvek reflecting area to the total inner surface to approximately 97.8\%. The green points in Fig. \ref{muon_onestep} represent the mean step of photons at different time ranges after the generation of photons, averaged with the read out values from the LHAASO-MD prototype simulation with a water absorption length of 100~m and a Tyvek reflectivity of 98.5\%. The black points represent the mean step length of photons in the Cherenkov tank with the same LHAASO-MD, but with a detector height of 90~cm, while the red points are the values for the tank with a height of 60~cm. It can be seen that the mean step length changes with the detector height, but remains stable when the Tyvek area is slightly changed by the vertical black strips.

Assume that a number $N_{0}$ of photons with water absorption length ($\lambda$) and Tyvek reflectivity ($f$) flying in the tank at a certain time when the mean step length $L$ becomes stable, while $r$ percent of the inner surface is covered by Tyvek and the reflection of the left part is 0. For another time $t$, the total transmission distance of each survival photons is $t$ multiplied by the velocity of photons. The number of photons with flying distance $x$ becomes
\begin{equation}
 N_{0} \times e^{-\frac{x}{\lambda}} \times f^{\frac{x}{L}} \times r^{\frac{x}{L}}.
\label{yx}
\end{equation}
The decreasing function is not affected by the scattering factor of the medium in the tank, because the scattered photons are still in the tank. For water and air, the scattering effect is very small compared to the water absorption and Tyvek absorption, while the effect of scattering on the mean step length is negligible as well.

In the meantime, the number of photons decreases with the function $N_{0} \times e^{-\frac{x}{\lambda_m}}$, where ${\lambda_m}$ is the effective attenuation length of photons in the tank determined by the total absorption effect in the tank. There is formula:
\begin{equation}
    e^{-\frac{x}{\lambda_m}} = e^{-\frac{x}{\lambda}} \times (f \times r)^{\frac{x}{L}}.
\label{eqn:reflectivityx}
\end{equation}

The probability of a photon hitting the PMT photocathode is constant for each reflection by the liner, and the decrease factor of photons hitting the PMT is equal to that of the total photons in the tank, when it contains enough photons. The number of total photons in the tank and as well as the photons hitting the PMT decrease with the transmission distance according to $e^{-\frac{x}{\lambda_m}}$ and decrease with the time according to $e^{-\frac{t}{\tau}}$. For single wave photons in the tank, the time distribution of photons hitting the PMT can be fitted to the function $N_{0} \times e^{-\frac{t}{\tau}}$, then the effective attenuation length ${\lambda_m}$ can be obtained by multiplying $\tau$ with the velocity of photons.

The formula \ref{eqn:reflectivityx} can be rewritten as

\begin{equation}
 \lambda_m =1/(\frac{1}{\lambda} -\frac{lnf}{L}- \frac{ln(r)}{L}).
\label{eqn:reflectivitc}
\end{equation}
Thus, the effective attenuation length of photons measured by the PMT is the function of the water absorption length $\lambda$, Tyvek reflectivity $f$, ratio of the Tyvek area to the total inner surface $r$, and the mean step length $L$ of photons in the tank. The effective attenuation length can be measured by using a LED providing pulse light with certain wavelength, and by measuring the reflection of photons in the tank with the PMT. For the cylindrical tank with a Tyvek inner surface, the mean step length is found to be constant when the ratio of the Tyvek area is changed from 100\% to 95\% by shielding the side Tyvek with black vertical strips of different widths (as shown in Fig.~\ref{mcsp_onestep}). After measuring the effective attenuation length of photons with different ratios of Tyvek areas, the mean step length can be fitted out with the formula in Eq. (\ref{eqn:reflectivitc}) by fitting the data points of $\lambda_m$ versus $r$. Simultaneously, the total effect of the value of $\frac{1}{\lambda} -\frac{lnf}{L}$ can also be fitted out with the function
\begin{equation}
 \lambda_m =1/(a- \frac{ln(r)}{L}).
\label{eqn:reflectivitd}
\end{equation}

Thus, the mean step length can be measured by changing the area of the Tyvek in the tank even if $\lambda$ and $f$ are unknown; however, $\lambda$ or $f$ can be determined if another value is known. If the medium in the tank is air, absorption can be ignored, hence
\begin{equation}
 \lambda_m =1/(-\frac{ln(f)}{L}- \frac{ln(r)}{L}).
\label{eqn:reflectivite}
\end{equation}
Thus, the mean step length of photons in the tank and the Tyvek reflectivity $f$ in air can be measured simultaneously.

For the water Cherenkov detector, the inner surface is usually completely covered by Tyvek, excluding the PMT area, and the PMT absorption can be ignored if the PMT area is small compared to the inner surface. For this case, Eq. (\ref{eqn:reflectivitc}) can be written as
\begin{equation}
 \lambda_m =1/(\frac{1}{\lambda} -\frac{lnf}{L}),
\label{eqn:reflectivita}
\end{equation}
where $\lambda_{m}$ is the attenuation length due to the Tyvek absorption and water absorption. The mean step length of photons in the water Cherenkov detector is determined by the geometry of detector, thus it changes with the height of detector (as shown in Fig.~\ref{mcsp_onestep}). The mean step length can be measured directly, or can be determined by simulation as well. Measuring the effective attenuation length of photons by the PMT with different heights of water, can provide data points of the effective attenuation length $\lambda$ versus the mean step length. Then, the water absorption length and Tyvek reflectivity in the water can be simultaneously obtained by fitting the data with Eq. (\ref{eqn:reflectivita}).

\section{Verification of the novel method for measuring the step length}
\subsection{Simulation}
\subsubsection{Simulation of the water condition}
\begin{figure}[!ht]
  \centering
  \includegraphics[width=3.4in]{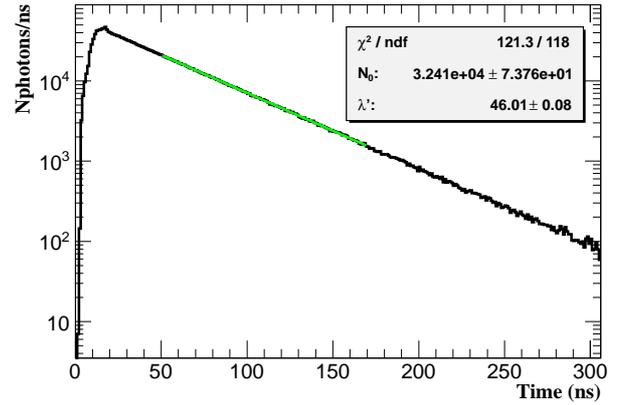}
  \caption{Time distribution of photons hitting the PMT, from the simulation of a tank with a diameter of 90~cm and height of 88~cm, and inputs $\lambda$=100~m and $f$=95\%.}
  \label{mcsp_time}
\end{figure}

Firstly, a GEANT4 simulation was performed to test the novel method of measuring step length. For the calculations, a cylinder tank with the radius of 90~cm and height of 88~cm was used, with Tyvek as inner reflecting surface, containing pure water. The water absorption length was set to 100~m and the Tyvek reflectivity was set to 95\%. A small hole with a diameter of 1.5~cm is opened on the top Tyvek, and a PMT is on the top facing downward to collect photons in the tank through the small hole. Initially all the inner surfaces, except the PMT area was set as Tyvek, and the parameters describing the Tyvek reflection model were the same as those used in the LHAASO-MD prototype simulation described above. In the simulation, 1 billion photons with a wave length of 405~nm were generated in 10~ns, and photons reaching the PMT were collected to obtain the time distribution of photons hitting the PMT (see Fig.~\ref{mcsp_time}). The step length of photons between each reflection can also be obtained by the simulation. The black points in Fig.~\ref{mcsp_onestep} are the step length of photons at different time after the generation of photons in the tank, and the red points are the mean values at different time ranges.

Then, a long vertical strip of the side Tyvek with a height of 80~cm and width of 2~cm was removed to achieve an area of 0 reflectivity, and photons were generated and simulated to obtain the time distribution of photons hitting the PMT under this condition. Then, the width of the dark surface was set to 4~cm, 6~cm, 8~cm, and 10~cm. The red points in Fig.~\ref{mcsp_onestep} represent the mean step length values at different time ranges with the width of the black strip of 10~cm. It can be seen that the step length becomes constant after 20~ns and the mean step length is similar for different ratios of Tyvek area under these conditions. In this process, the ratio of the Tyvek area to the inner surface was calculated by considering the area of the dark area and the PMT area, and the calculated ratio of the Tyvek area was ranging from 99.986\% to 97.808\%.

\begin{figure}[!ht]
  \centering
  \includegraphics[width=3.4in]{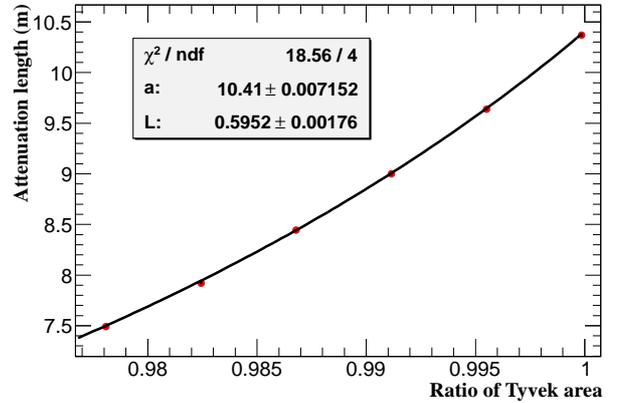}
  \caption{Attenuation length versus ratio of Tyvek area obtained from the simulation of a tank with a diameter of 90~cm and height of 88~cm, and inputs $\lambda$=100~m and $f$=95\%.}
  \label{mcsp_len_ratio}
\end{figure}

The time distribution of photons hitting the PMT was determined from each simulation, fitted with the function $N_{0} \times e^{-\frac{t}{\tau}}$, and the attenuation length was obtained by multiplying $\tau$ with the velocity of photon. The time range with exponential distribution and at least 1000 photons hitting PMT in each ns is the reasonable fitting range, with constant mean step length in the range. The time distribution from each condition was fitted in three differ ranges(30-140 ns, 50-170 ns, 70-190 ns), and three attenuation length were obtained for the systematic error calculation considering the fitting error caused by statistics is negligible. Figure~\ref{mcsp_len_ratio} shows the attenuation length versus the ratio of Tyvek area obtained from the simulation in each condition. The y-axis of each red point in Fig.~\ref{mcsp_len_ratio} is the mean value of the attenuation length given by the time distribution fitted in three different ranges, and the error bar shows the largest bias of the mean value for the three results. The black line is the fitting with the function in Eq. (\ref{eqn:reflectivitd}), with the mean step length obtained from the fitting is approximately 0.5932~m, which is in good agreement with the value about 0.6~m averaged with the read out values from the simulation shown in Fig~\ref{mcsp_onestep}. The total effect of $\frac{1}{\lambda} -\frac{lnf}{L}$ is also determined from the fitting.

\subsubsection{Simulation of the air condition}
\begin{figure}[!ht]
  \centering
  \includegraphics[width=3.4in]{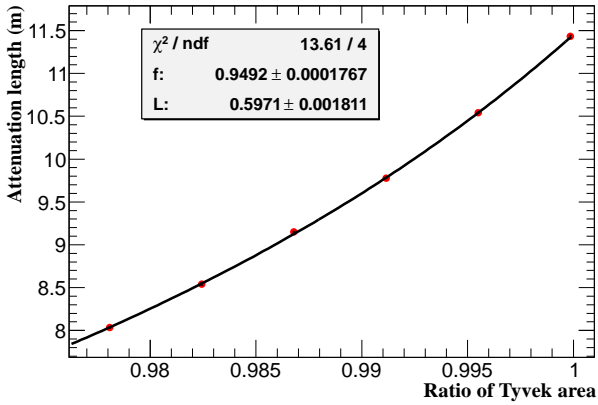}
  \caption{Attenuation length versus ratio of the Tyvek area determined by the simulation of the tank with a diameter of 90~cm and height of 88~cm, and inputs $\lambda$=1000~m and $f$=95\%.}
  \label{mcair_attrate}
\end{figure}

 If the absorption length ${\lambda}$ is sufficiently high, the absorption of the medium can be ignored, and the step length and Tyvek reflectivity can be measured simultaneously by changing the inner Tyvek area of the cylindrical tank. Another simulation was performed under the same conditions to the process above, but the absorption length was set to 1000~m. The tank diameter and height remained 90~cm and 88 cm, respectively, and the side Tyvek was shielded with a different vertical black strips with a height of 80~cm and widths in the range of 0--10~cm. Photons were generated for each simulation and the time distribution of photons hitting the PMT was determined and fitted in three different ranges to obtain the attenuation length. Figure~\ref{mcair_attrate} shows the effective attenuation length versus the ratio of Tyvek area from the simulation with an absorption length of 1000~m and Tyvek reflectivity of 95\%. The results from the fit with the function $\lambda_m =1/(-\frac{ln(f)}{L}- \frac{ln(r)}{L}$) are shown in Fig.~\ref{mcair_attrate}.
The reflectivity is 94.93\% and mean step length is 0.5965~m, which are in very good agreement with the input and expected values, with a bias for the reflectivity less than 0.1\% and the bias for a mean step length less than 1\%. The absorption length of air is usually greater than 1000~m, thus, the step length and the Tyvek reflectivity in the air can be measured simultaneously with this method.

\subsection{Experiment for the measurement of step length and reflectivity in air}
\begin{figure}[!ht]
  \centering
  \includegraphics[width=3.4in]{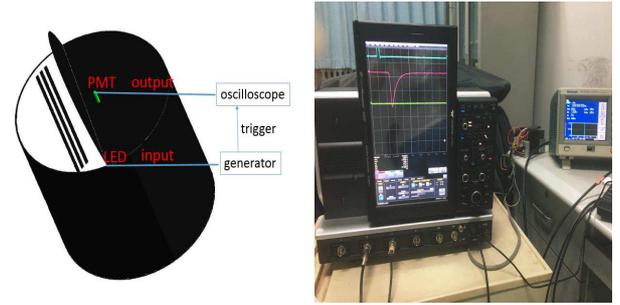}
  \caption{Left pane: schematic diagram of the experiment. Right pane: image of the real experimental and data acquisition system}
  \label{exp}
\end{figure}

\begin{figure}[!ht]
  \centering
  \includegraphics[width=3.4in]{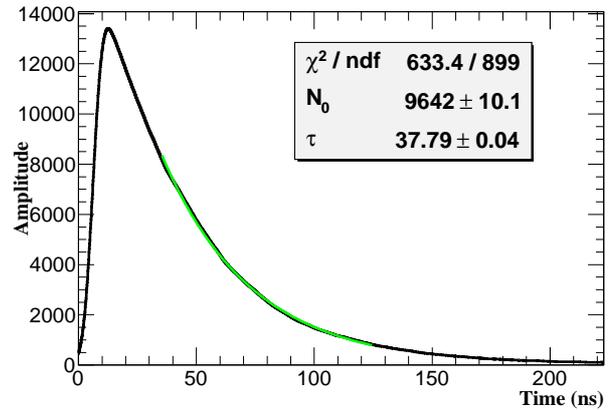}
  \caption{Time distribution of photons hitting the PMT from the experiment for measuring the step length and reflectivity}
  \label{dataair_time}
\end{figure}

A cylindrical hollow shell tank with 90-cm diameter and 88-cm height was made to measure the step length and Tyvek reflectivity in air, using Tyvek as the inner surface and the tank filled with air. The left pane in Fig.~\ref{exp} shows a schematic diagram of the experiment, and the right pane shows an image of the real experimental and data acquisition system. The tank was shielded by a dark sheet with no light leakage, and a LED generating photons with a wavelength of 405~nm was placed in it. Firstly, the inner surface is completely covered by Tyvek 1082D, except for a 1.5-cm diameter round hole on the top for the PMT to collect photons. The LED was driven by a pulse generator with 10~ns width and 1~kHz frequency, and the photons hitting the PMT were collected by an oscilloscope to obtain the waveform of the time distribution of photons. Data acquisition was trigged by another signal from the generator with the same start time and frequency as that of the signal to drive LED. Figure~\ref{dataair_time} shows the averaged waveform of the PMT signals of more than 1000 events, which represents the time distribution of photons hitting the PMT. The PMT waveform was fitted with the function $N_{0} \times e^{-\frac{t}{\tau}}$, then the time decay factor was converted to effective attenuation length by multiplying the velocity of light.

\begin{figure}[!ht]
  \centering
  \includegraphics[width=3.4in]{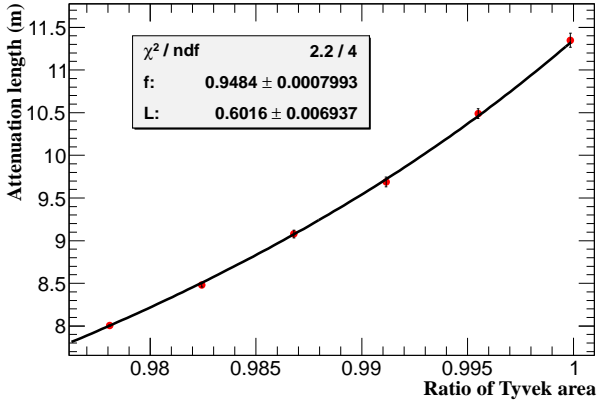}
  \caption{Attenuation length versus ratio of the Tyvek area from the experiment for measuring the step length and reflectivity}
  \label{dataair_len_ratio}
\end{figure}

Then, the side Tyvek is shield with an 80-cm high vertical black strip with widths of 2~cm, 4~cm, 6~cm, 8~cm, and 10~cm. The LED was driven by a pulse generator and the PMT waveform was obtained by the PMT under each condition. The effective attenuation length was determined from the fitting under each condition in three different ranges, and the ratio of the Tyvek area was calculated. The red points in Fig.~\ref{dataair_len_ratio} represents the effective attenuation length versus the ratio of the Tyvek. Each effective attenuation length is the mean value from the relative PMT waveform fitted in three different ranges, and the error is the largest bias with the mean value. The black line represents the fitting with function $\lambda_m =1/(-\frac{ln(f)}{L}- \frac{ln(r)}{L}$), and the mean step length and Tyvek reflectivity was determined from the fitting simultaneously. The measured reflectivity is 94.84\% and the step length is 0.6016~m with some errors. The step length is in good agreement with the results from the simulation. The measured reflectivity is consistent with the Tyvek reflectivity measured by the Bureau of Measures in the error range. The measured reflectivity was the total reflectivity of the inner surface, which had contributions other than that of Tyvek. If the outside of Tyvek was directly shielded by the black sheet, eliminating the effect from the outer steel, then the measured reflectivity equals to the sole reflectivity of the Tyvek.

The error due to the size of the detector is less than 0.15\%, which was obtained by fitting the data points of attenuation length versus ratio of Tyvek area, with ratios of Tyvek area calculated by changing the diameter and height of the tank with 1 cm. The error due to the reflection of the black sheet is less than 0.1\%, determined by supposing the reflectivity of the dark sheet as 0.1\%. Typically, the error caused by the PMT performance can be ignored for the fitting of PMT waveform if the statistic is high enough, as the sigma of single photon electron is usually less than 2 ns. The time range with exponential distribution and high statistics for the PMT waveform is the reasonable fitting range, without the first part affected by the initial directions of photons and the tail of the waveform with low statistics. The error due to the fitting of the PMT waveform were calculated with the fitting in three different ranges, which is less than 0.1\% for reflectivity and less than 1.2\% for mean step length (see Fig.~\ref{dataair_len_ratio}).

\section{A novel method for the simultaneous measurement of reflectivity and water absorption length }
\subsection{Test with a small tank}
\begin{figure}[!ht]
  \centering
  \includegraphics[width=3.4in]{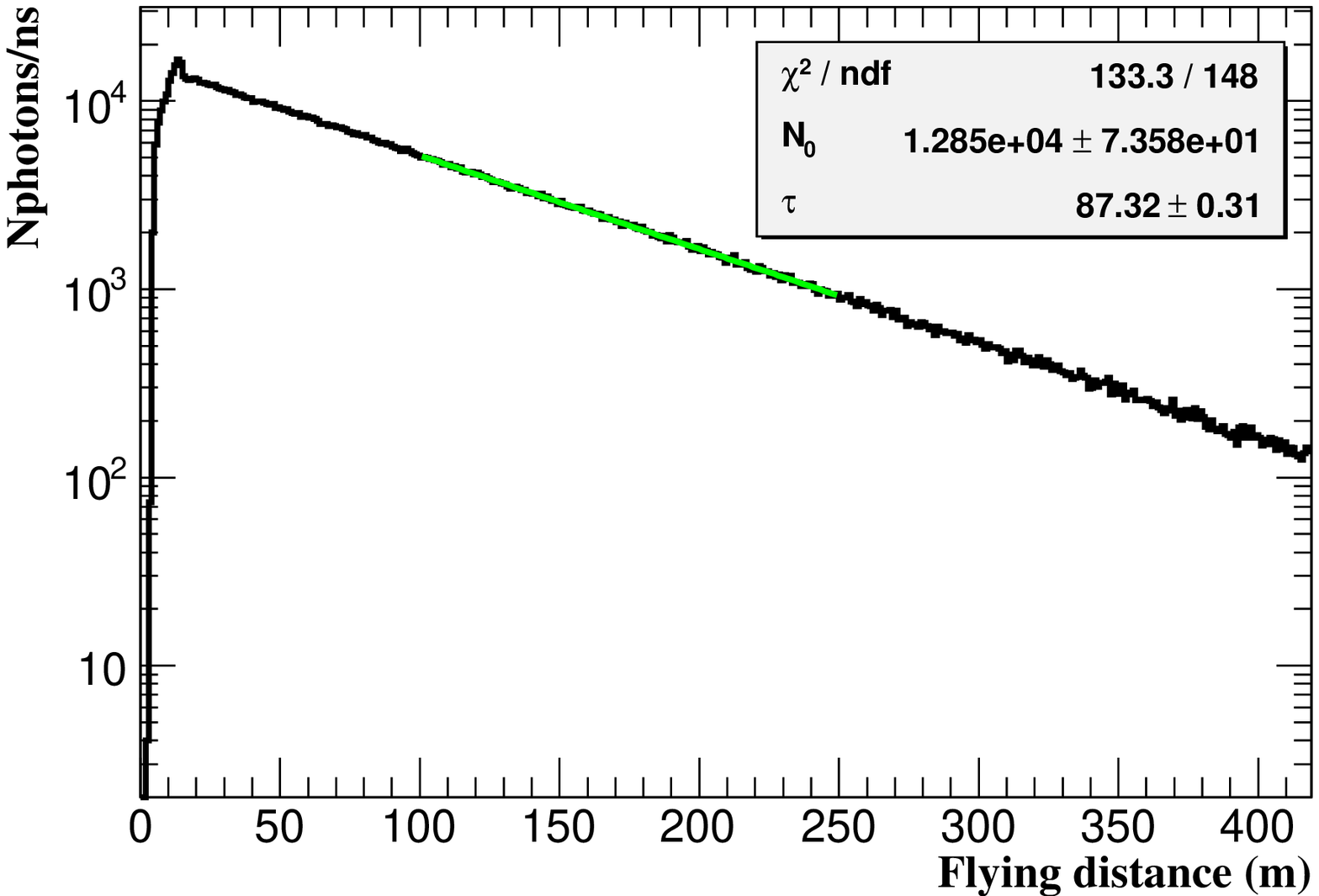}
  \caption{Time distribution of photons hitting the PMT from the simulation of the tank with a diameter of 100~cm and height of 70~cm, and inputs $\lambda$=100m and $f$=98\%}
  \label{water2_wave}
\end{figure}

\begin{figure}[!ht]
  \centering
  \includegraphics[width=3.4in]{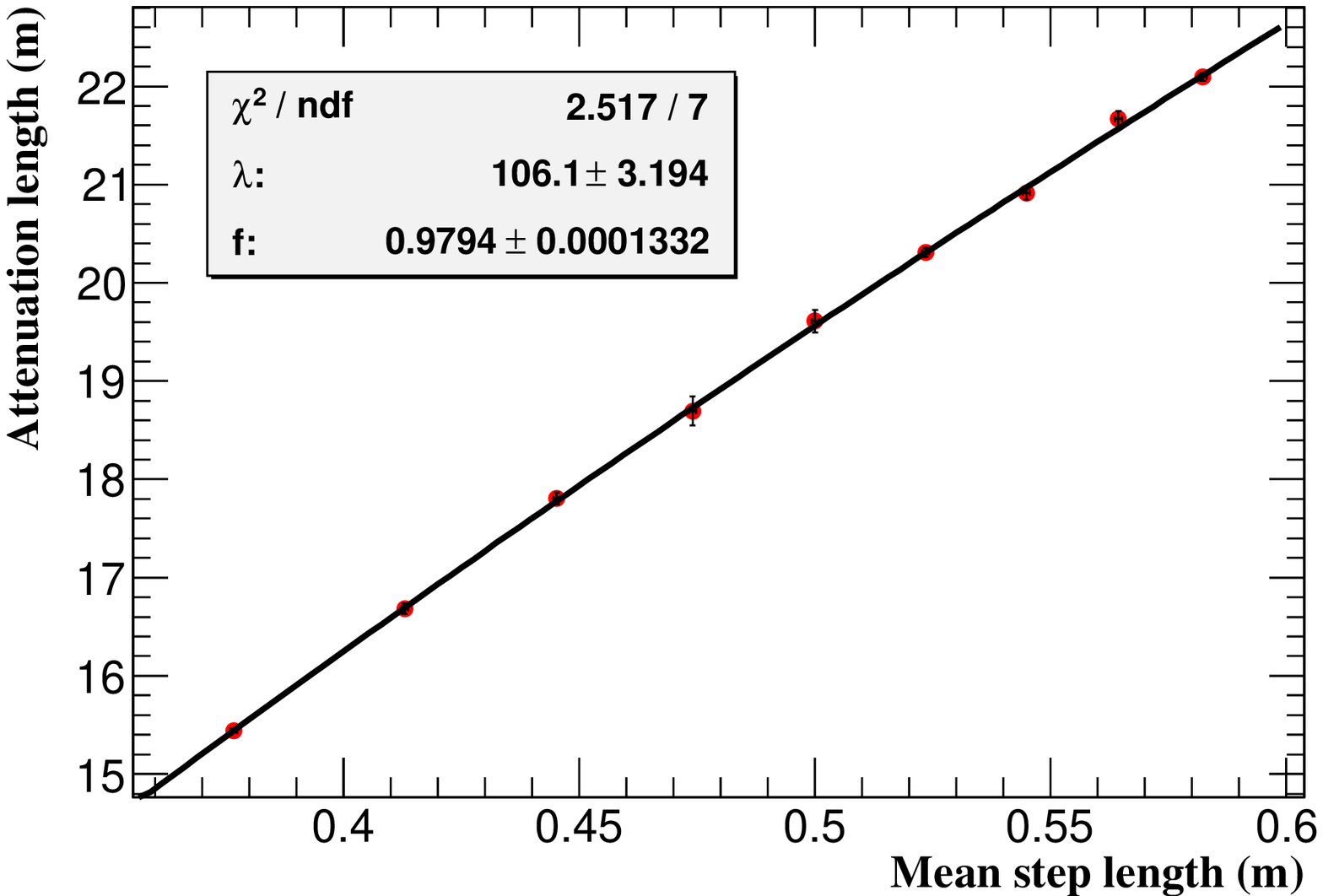}
  \caption{Attenuation length versus mean step length from the simulation of the tank with a diameter of 100~cm and different heights, and inputs $\lambda$=100m and $f$=98\%}
  \label{water2_attlen_slen}
\end{figure}

Firstly, a cylinder tank with a diameter of 1~m was used in the simulation test. A small PMT was placed on the top of the Tyvek, allowing it to move with the height of water. The height of the detector was changed from 30~cm to 80~cm in steps of 5~cm for simulation. For each detector height, the total inner surface of the detector was covered by Tyvek, except for a 1.5-cm diameter hole on its top for the PMT. Photons with a wavelength of 405~nm were generated in 10~ns for each detector height, and the time distribution of photons hitting the PMT was determined (see Fig.~\ref{water2_wave}). The Tyvek reflectivity was set to 98\%, the water absorption length was set to 100~m, and the parameters for the Tyvek model were same as those for the LHAASO-MD prototype simulation. The mean step length was determined from the simulation for each detector height, with the total transmission distance divided by the number of reflections for photons with constant mean step length. The effective attenuation length was obtained by fitting the PMT time distribution with the function $N_{0} \times e^{-\frac{t}{\tau}}$ (see Fig.~\ref{water2_wave}), then effective attenuation length were obtained. The red points in Fig.~\ref{water2_attlen_slen} represent the effective attenuation length versus the mean step length. Each effective attenuation length is the mean value from the time distribution fitted in three different ranges, and the error is the largest bias with the mean value. The black line is the fitting with Eq. (\ref{eqn:reflectivityx}). The result from the fitting for the Tyvek reflectivity is 97.94\% and for the water absorption length is 106.1~m, with an error less than 0.1\% for reflectivity and an error less than 10\% for the water absorption length. The errors can be further reduced by considering the PMT absorption for the fitting of the time distribution with the function $N_{0} \times e^{-\frac{t}{\tau}} \times (1-p)^{\frac{t \times v}{L}}$, where p is the ratio of PMT area to the total inner surface area and v is the velocity of photons.
An experiment about the same process was done, with the measured Tyvek reflectivity and water absorbing length in the expected regions.

\subsection{Expectation for the LHAASO-MD parameter measurement}
Each LHAASO-MD is a water Cherenkov muon detector with a diameter of 6.8~m and height of 1.2~m, with an 8-inch PMT mounted on the top facing downward. The optical performance of Tyvek is expected to be similar for different purity conditions of water in the tank, which can be measured in advance. The water absorption length depends on the quality and purity of the water, thus for the MD performance study it is advisable to measure the water absorption length of water in the MD tanks. Each MD has a Tyvek bag containing water, and the height of the top of the Tyvek changes with the height of the water during the installation the MD when filling it with water. The PMT waveform of muon signals and signals of LED photons with certain wavelength can be measured with different heights of water while filling water to the MDs.
The attenuation length of photons can be determined from the fitting of the PMT waveform of the LED signals, and the step length can be obtained from simulation or can be measured in advance. Thus, the water absorption length and Tyvek reflectivity can be determined from fitting.

\begin{figure}[!ht]
  \centering
  \includegraphics[width=3.4in]{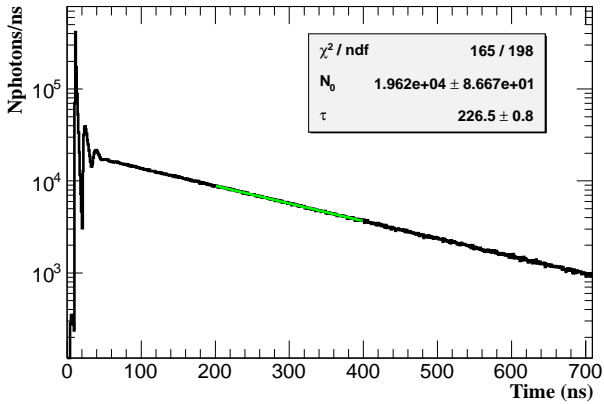}
  \caption{Time distribution of photons hitting the PMT from the simulation of the LHAASO-MD tank with a diameter of 6.8~m and height of 1.2~m, and inputs $\lambda$=100~m and $f$=98.5\%}
  \label{muon_wave}
\end{figure}

\begin{figure}[!ht]
  \centering
  \includegraphics[width=3.4in]{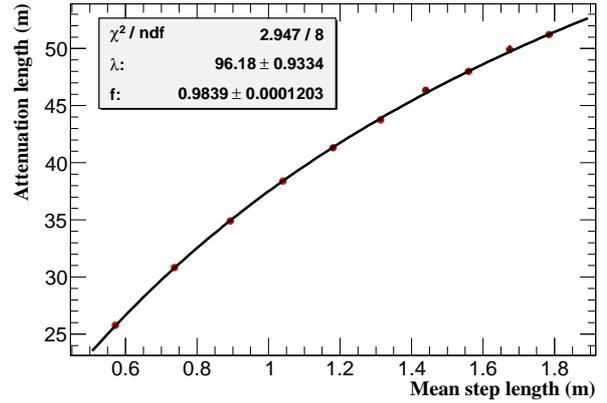}
  \caption{Attenuation length versus mean step length from the simulation of the LHAASO-MD tank with a diameter of 6.8~m and different heights, and inputs $\lambda$=100~m and $f$=98.5\%}
  \label{muon_attlen_slen}
\end{figure}

A simulation of this process was performed with the MD detector, with a tank diameter of 1.2~m and with an 8-inch PMT mounted on the top, by changing the height of water from 30~cm to 120~cm with a step of 10~cm. The Tyvek reflectivity was set to 98.5\%, the water absorption length was set to 100~m in the simulation. Photons with 405~nm wavelength were generated for each simulation, and the time distribution of photons hitting the PMT and the mean step length for each water height were determined from the simulation. Figure~\ref{muon_wave} shows the time distribution of photons at the water height of 120~cm, and the fitting with the function $N_{0} \times e^{-\frac{t}{\tau}}$ to get the effective attenuation length. The red points in Fig.~\ref{muon_attlen_slen} are the effective attenuation length versus the mean step length. Each effective attenuation length is the mean value from the time distribution fitted in three different ranges, and the error is the largest bias with the mean value. The black line in Fig.~\ref{muon_attlen_slen} represents the fitting with Eq. (\ref{eqn:reflectivityx}). The result from the fitting for the Tyvek reflectivity is 98.39\% and for the water absorption length is 96.18~m, with an error less than 0.13\% for reflectivity and a bias less than 5\% for the water absorption length. The errors can be further reduced by considering the PMT absorption for the fitting of the time distribution.

\section{Conclusion and discussion}
We have developed novel methods for the measurement of water absorption length, Tyvek reflectivity, and mean step length of photons. These can be used for water Cherenkov experiment, and can be used in studies in optical physics as well. For example, the relationship of the water absorption length and Tyvek reflectivity between wavelength of photons and purity of water can be measured accurately. The parameters for the Tyvek reflection model can be fine-tuned after the mean step length is measured. Apart from Tyvek, the reflectivity of other high-reflectivity materials with diffuse reflection as the main part can also be measured.

\section*{Acknowledgements}
This work was supported by the National Natural Science Foundation of China (Nos. 11405181 and 11375210).
We gratefully acknowledge the assistant and helpful discussion from Professor Zhen Cao, Liming Song, Changzheng Yuan, Xiangdong Sheng, Min Zha and Yi Zhang. Thanks also go to
all the people who gave help and supports for this work.


\section*{References}
\begin{thebibliography}{00}

 \bibitem{auger} A. Aab et al., The Pierre Auger Collaboration, The Pierre Auger Cosmic Ray Observatory, Nuclear Inst. and Methods in Physics Research A 798 (2015) p. 172-213
 \bibitem{dayabay} Z. Yu et al., Study of a prototype water Cherenkov detector for the DayaBay neutrino experiment, Nuclear Inst. and Methods in Physics Research A 682 (2012) p. 26每30.
 \bibitem{superk} K. Abe et al., Calibration of the Super-Kamiokande detector, Nuclear Inst. and Methods in Physics Research A 737 (2014) p. 253每272
 \bibitem{asgam} C. Liu et al., Performance of the muon detector A under TIBET III array, Chinese Physics C. Vol.37, No.2 (2013) 026001
 \bibitem{CAOZ_2014nim} Z. Cao, LHAASO Collaboration, Status of LHAASO updates from ARGO-YBJ, Nuclear Instruments and Methods in Physics Research A 742 (2014) 95每98
 \bibitem{CAOZ_2014} Z. Cao., LHAASO Collaboration, LHAASO: Science and Status, Frascati Physics Series Vol. 58(2014). p. 331-342
 \bibitem{ZUO_2015} X. Zuo, LHAASO Collaboration, Design and performances of prototype muon detectors of LHAASO-KM2A, Nuclear Inst. and Methods in Physics Research A 789 (2015) p. 143-149.
 \bibitem{lixr_2015} XR. Li and HH. He et.al, Study of Water Cherenkov muon detector parameters and performances for LHAASO, Proc. of Sci. 0871, 34st ICRC (2015).
 \bibitem{length1} R. C. Smith and K. S. Baker, Optical properties of the clearest natural waters (200--800 nm), Applied Optics, Vol. 20, No. 2, (1981) p. 177-184
 \bibitem{length2} MJ. Chen and YF. Wang et al., A water Tank Prototype for the Cerenkov Calorimeter, Chinese Physics C, Vol.29, No.10 (2005) p. 983-988
 \bibitem{length3} F. Amat et al., Measuring the attenuation length of water in the CHIPS-M water Cherenkov detector, Nuclear Inst. and Methods in Physics Research A 844 (2017) 108每115
 \bibitem{length4} F. M. Sogandares and E. S. Fry, Absorption spectrum (340--640) nm of pure water. I. Photothermal measurements, Applied Optics, Vol. 36, No. 33 (1997) p. 8699-8709
 \bibitem{length5} R. M. Pope and E. S. Fry, Absorption spectrum (380--700) nm of pure water. II. Integrating cavity measurements, Applied Optics, Vol. 36, No. 33 (1997) p. 8710-8723
 \bibitem{angle1} A. Filevich et al., Spectral-directional reflectivity of Tyvek immersed in water, Nuclear Instruments and Methods in Physics Research A, Vol. 423, No. 1 (1999) p. 108-118
 \bibitem{angle3} J.C. Arteaga Vel\'{a}zquez et al., A measurement of the diffuse reflectivity of 1056 Tyvek in air and water, Nuclear Instruments and Methods in Physics Research A Vol. 553, No. 1-2, (2005) p. 312-316
\bibitem{wly} LY. Wang et al., Study of Tyvek rectivity in water, Chinese Physics C, Vol. 36, No. 7 (2012) 628-632
\bibitem{cavity1} Dana Z. Anderson et al., Mirror reflectometer based on optical cavity decay time, APPLIED OPTICS Vol. 23, No. 8 (1984) p. 1238-1245
\bibitem{cavity2} Li-feng GAO et al., High reflectivity measurement with cavity ring-down technique, Proc. of SPIE Vol. 5963 (2005) 59632F-1


\end{thebibliography}
\end{document}